# Polarized neutron channeling in weakly magnetic films

S.V. Kozhevnikov[1]\*, V.D. Zhaketov[1], T. Keller[2,3], Yu.N. Khaydukov[2,3], F. Ott[4], F. Radu[5]

[1]*Frank Laboratory of Neutron Physics, JINR, 141980 Dubna Moscow Region, Russia*
*\*e-mail: kozhevn@nf.jinr.ru*
[2]*Max Planck Institut für Festkörperforschung, Heisenbergstr. 1, D-70569 Stuttgart, Germany*
[3]*Max Planck Society Outstation at FRM-II, D-85747 Garching, Germany*
[4]*Laboratoire Léon Brillouin CEA/CNRS, IRAMIS, Université Paris-Saclay, F-91191 Gif sur Yvette, France*
[5]*Helmholtz-Zentrum Berlin für Materialien und Energie, Albert-Einstein Straße 15, D-12489 Berlin, Germany*

We describe a new sensitive method for the investigation of weakly magnetic films placed inside a tri-layer planar waveguide. Polarized neutrons tunnel into the waveguide through the surface, channel along the layers and are emitted from the end face as a narrow and slightly divergent microbeam. Polarization analysis permits to detect very small magnetization in the order of a few 10 Gauss. The magnetic film containing the rare-earth element Tb was investigated using both fixed wavelength and time-of-flight polarized neutron reflectometers. The experimental results are presented and discussed.

## I. INTRODUCTION

Thin magnetic films are widely used for practical applications and fundamental investigations. Therefore the development of new methods for its characterization is a timely task. A powerful tool for the investigation of thin magnetic films is Polarized Neutron Reflectometry (PNR) [1-5] which allows to extract the Scattering Length Density (SLD) for neutron spin up (+) and down (-) as a function of the coordinate $z$ perpendicular to the film surface

$$\rho^{\pm}(z) = \rho_N(z) \pm cB \quad (1)$$

where $\rho_N(z)$ is the nuclear SLD, c is a constant and $B$ is a magnetic induction. The direction and the magnitude of the magnetization vector can be extracted for each layer of a multilayer film from a fit of model calculations to experimental reflectivities. The drawback of PNR is that low magnetization values can be hardly resolved. The smallest magnetization detected by conventional reflectometry is about 1000 G. To extract a low magnetization value about 100 G we need a typical measuring time about 100 hours, too much for conventional experiments. Nevertheless, in the particular case of Bragg diffraction in thin films it was possible to extract a low magnetization about 10 G [6].

Thin magnetic films containing rare-earth elements are promising materials for the development of new methods of magnetic data storage and ultra-fast switching [7], and have a low magnetization inaccessible for PNR. Recently we demonstrated the new method of Polarized Neutron Channeling (PNC) in planar waveguides to extract magnetization value of a few 10 G [8,9] in $TbCo_5$ films with saturation magnetization around 200 G [10]. In this work we present new results on the $TbCo_{11}$ films with the higher saturation magnetization, and compare the performance of fixed wavelength and the time-of-flight polarized neutron reflectometers for PNC studies.

Various layered resonator systems might be used for the investigations of weakly magnetic films: i) interference filters; ii) Fabry-Perot resonators and iii) resonators or planar waveguides. All of them have a very similar potential well structure but slightly different features and applications. We give a short review of literature to identify adequate resonator systems for the determination of small magnetization.

In Fig. 1a the experimental geometry for *interference filter* is shown. Neutron beam in air or vacuum (medium 0) irradiates the surface of a tri-layer system under the grazing angle $\alpha_i$. The SLD has a shape of a potential well (Fig. 1b) where the middle layer (medium 2) with low SLD is sandwiched by two layers with high SLD (media 1 and 3). Neutrons tunnel through the upper thin layer. In the middle layer the neutron wavefunction density is resonantly enhanced. Neutrons tunnel through the bottom thin layer and refracted in a substrate. If the middle layer is thin, a pure resonant minimum at total reflection (Fig. 1c) and corresponding maximum in transmission (Fig. 1d) arise. This phenomenon is termed as *frustrated total reflection*. The SLD wells correspond to the energy of ultracold neutrons around 100 neV. Therefore interference filters were widely used for monochromatization and spectrometry of ultracold neutrons. The first multilayer interference filter was proposed in 1974 [11] and the first experiments with neutron interference filters were described in [12,13]. The review on fundamental experiments with ultracold neutrons using interference filters was done in [14].

If the middle layer is relatively thick as in Fig. 2a, then we observe many deep resonances in the region of total reflection (Fig. 2b) and corresponding maxima in transmission (Fig. 2c). Such a tri-layer structure is the neutron analog of a *Fabry-Perot interferometer* [15]. In review [16] these sensitive devices are proposed for fundamental investigations of the neutron life-time and Goos-Hänchen effect, for physical investigations of surface magnetism and surface superconductivity, and also for neutron optical devices, such as monochromators, polarizers, beam-splitters, and interferometers.

The third tri-layer resonator system has a thin upper layer, a middle layer of the thickness $d$, and a thick bottom layer. The geometry is shown in Fig. 3a and the SLD is presented in Fig. 3b. The neutron wave tunnels through the upper thin layer into the middle layer, is almost totally reflected from the bottom thick layer, partially exits through the upper layer in the direction of specularly reflected beam (marked as $R_{ch}$), and is partially reflected from the upper layer back to the middle layer.

In the region of total reflection the reflectivity has weak dips (Fig. 3c) due to resonant enhancement of the neutron wavefunction density in the middle layer (Fig. 3d). Then neutrons propagate along the middle layer as in a channel and finally are emitted from the end layer as a narrow but slightly divergent microbeam. The divergence of this microbeam is mainly defined by Fraunhofer diffraction on a narrow slit as $\sim \lambda/d$ where $\lambda$ is neutron wavelength. Neutrons partially exit through the upper layer and therefore leaks from the microbeam, the neutron wave function density decays exponentially along the guiding layer. This decay parameter is termed as neutron channeling length $x_e$ and typically consists of several millimeters. When the resonant enhancement of the neutron wavefunction density is used then this structure is termed as *resonator*. If the channeling phenomenon is used then this system is called as *planar waveguide.*

For better understanding the difference between resonators and waveguides we review applications of resonators. The theory of neutron resonances in layered structures was developed in [17]. The resonances can be registered by various ways. The interaction of neutrons with matter leads to leakage of neutrons from the specularly reflected beam. On the total reflection plateau one can see dips corresponding to the resonances (Fig. 3c). It is the primary neutron channel for the registration of resonances. The secondary neutron channel is registration of resonant maxima for off-specular neutron scattering, arising from interface roughness, incoherent neutron scattering at hydrogen, spin-flip scattering in magnetically non-collinear layers and neutron channeling. The third channel is detection of secondary radiation, including gamma-rays, alpha-particles, protons, tritons, fission products resulting from nuclear reaction of neutrons with specific elements like as Gd, Li, U. Layered resonators were used for enhancement of weak neutron interaction with matter.

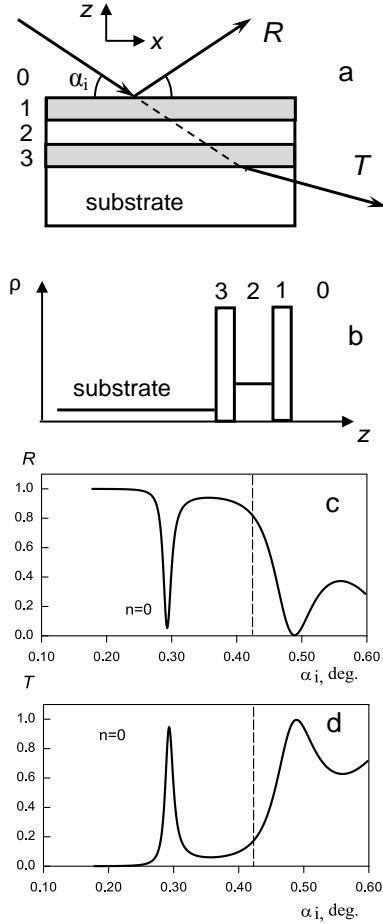

Fig. 1. Interference filter: (a) geometry; (b) SLD; (c) reflectivity; (d) transmission coefficient.

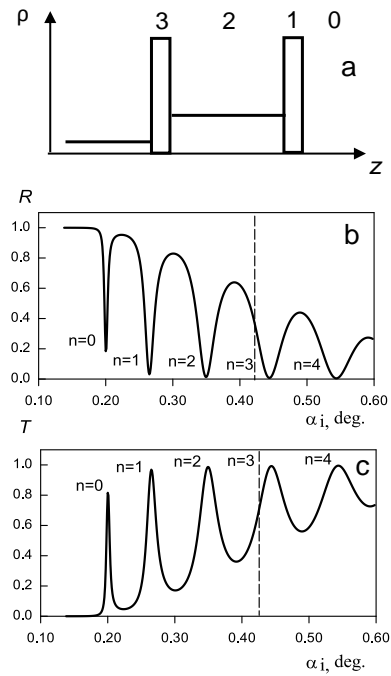

Fig. 2. Fabry-Perot interferometer: (a) SLD; (b) reflectivity; (c) transmission coefficient.

The dips on total neutron reflection from the layered polymer structure were registered in [18]. It was caused by incoherent neutron scattering from hydrogen. In [19] the dips on total neutron reflection and maxima of gamma-irradiation were observed for the layer $Gd_2O_3$. In [20] the resonant minima at total neutron reflection and resonant maxima of alpha-particles intensity were registered for the layer $^6LiF$. The enhanced off-specular neutron scattering due to interface roughness was observed in [21,22]. In [23-25] off-specular scattering of polarized neutrons was registered for the domain structure close interfaces. The spin-flipped neutron intensity was observed in [26] for the thin magnetic Co layer placed inside a resonator in a magnetic field applied under an angle to the sample surface. In [21] the spin-flipped intensity in specular reflection was registered for the resonator with a nonmagnetic middle layer and magnetic external layers. In the demagnetized state in a low parallel magnetic field the spin-flip reflectivity was registered. The review of methods of registration and application of neutron standing waves in layered structure can be found in [27]. Recently, the interest to use the layered resonator for the investigations is increased again. In [28] a layered resonator was used for the investigations of coexistence of magnetism and superconductivity. In [29,30] the polarized neutron beam was used to change the potential well structure and thus select the different layers for enhanced neutron interaction. In [31] it was proposed the magnetic layered resonator with uranium inside to create a compact atomic electrical power station. Using applied magnetic field it is possible to change the magnetization of the external magnetic layers and the potential well depth. It changes the coefficient of enhancement of the neutron wavefunction density inside the resonator with uranium what leads to reactivity modulation for the reaction of fission. In [30] the resonant maxima of incoherent neutron scattering from hydrogen containing layer were registered directly.

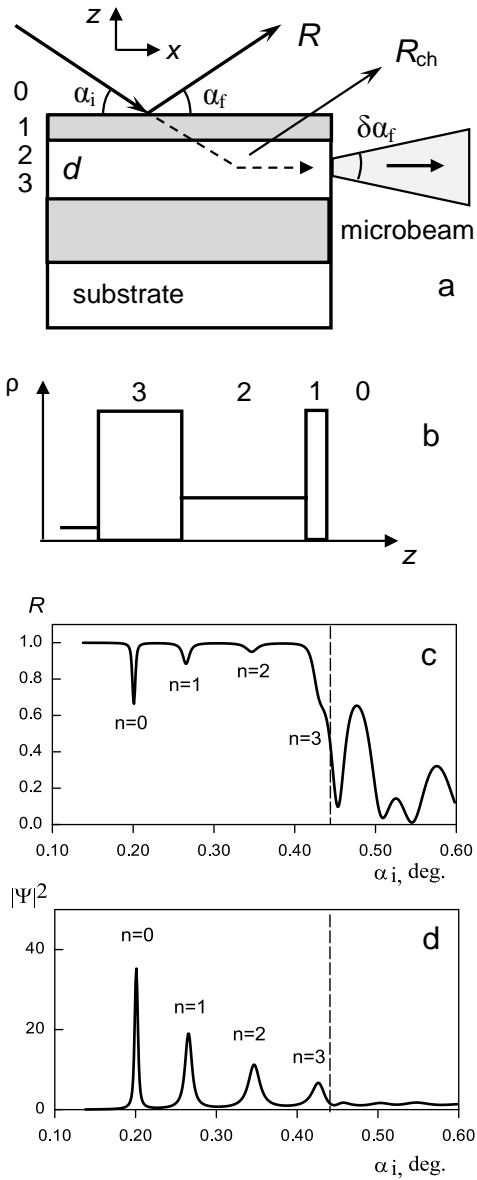

Fig. 3. Planar waveguide: (a) geometry; (b) SLD; (c) reflectivity; (d) neutron wavefunction density.

A planar waveguide is used for the production of a neutron microbeam emitting from the end face or for neutron channeling in the middle guiding layer. The planar waveguide for producing the neutron microbeam was considered theoretically for the first time in [32]. Such complicate type of the waveguide is called 'prism-like waveguide' because of the principle of the neutron beam introduction into the guiding layer is similar to a refractive prism used in optics. The first part is the resonant beam coupler with a thin upper layer. The second part is the waveguide with thick upper layer. For the first time the neutron beam from the exit end face of the prism-like waveguide was observed in [33]. The neutron channeling was registered for the first time in [34] in the prism-like waveguide in the geometry of reflection. The waveguide had three parts: resonant beam-coupler, waveguide and resonant decoupler (the same as the first part). The prism-like waveguides have a rather complicate structure and therefore were not used broadly. The simple waveguide based on tri-layer structure as in Fig. 3a is more simple and effective device. The neutron channeling in the simple waveguide was observed for the first time in [35] in the geometry of reflection. From the end face of the simple waveguide was obtained unpolarized [36,37] and polarized [38] neutron microbeam. The polarizing magnetic waveguide Fe/Co/Fe was investigated in [39]. The theory of neutron channeling in planar waveguides was developed [40]. The channeling length was experimentally measured in [41-43] and investigated in [44,45]. The experimental setup and different ways of the channeling length measurement are described in [46].

In conventional experiment the width of the neutron beam is from 0.1 to 10 mm. For the investigations of local microstructures in bulk with high spatial resolution a very narrow neutron beam is needed. Therefore various focusing devices including Fresnel lenses, capillary lenses, elliptical neutron guides and bent crystal monochromators are developed [47]. But these devices have restriction due to physical properties of used materials or technology its treatment. Therefore the minimal achievable width of the focused beam is 50 μm. Neutron planar waveguides are more efficient focusing devices which produce a neutron microbeam of the width from 0.1 to 10 μm. In [48] the polarized neutron microbeam was used for the spatial scan of an amorphous magnetic microwire. The combination of a nonmagnetic waveguide and polarized neutron reflectometer was used [49]. The divergence of the microbeam for the neutron wavelength about 4 Å and the guiding layer thickness 150 nm is about 0.1°, it leads to the microbeam broadening of about 2 μm per mm distance from the exit. The Fraunhofer diffraction contribution $\sim \lambda / d$ into the microbeam angular divergence was investigated experimentally. The dependence $\sim \lambda$ was measured experimentally in [50,51] and $\sim 1/d$ in [51,52]. The intrinsic spectral width of the neutron resonances in the microbeam was estimated experimentally [53]. In [54] various methods of a neutron microbeam shaping are discussed: slits from absorbing materials producing the microbeam of the width of 50 μm, total reflection from a small size substrate producing the microbeam of the width of 20 μm and planar waveguides with the microbeam of the width of 2 μm. The most versatile method with high intensity is total reflection from the substrate but neutron waveguides produce a narrowest microbeam.

We have reviewed various tri-layered resonant systems. First, the method of neutron interferometry based on Fabry-Perot interferometer (Fig. 2) and proposed in [16] is complicated for practical realization and data interpretation. Therefore it is difficult to use this structure for the direct measurement of low magnetization. Second, the position of resonances strongly depends on SLD and a thickness of the middle layer and weakly depends on SLD and a thickness of the outer layers 1 and 3. If the layers 1 and 3 are magnetic then the magnetization value defines the resonance position by the indirect way through the reflection coefficients from the middle layer to outside which must be calculated using model. Therefore the investigated magnetic film should be placed as the middle layer. Thus, the interference filter (Fig. 1) and the planar waveguide (Fig. 3) with the weakly magnetic middle layer are appropriate for the direct determination of low magnetization. The idea to use polarized neutron channeling in planar waveguides for the precise determination of the magnetization was described in [55]. But calculations were done for more complicate waveguides of prism-like type in reflection geometry. The planar waveguide of simple type (Fig. 3) is much easier for practical realization and data treatment.

In Section 2 we consider the principles of magnetic planar waveguides and present some calculations. In Section 3 the experimental results in the fixed wavelength measuring mode are presented. In Section 4 the experimental results the time-of-flight techniques are shown. In Section 5 the results are discussed.

## II. MAGNETIC PLANAR WAVEGUIDE

The geometry of experiment is shown in Fig. 3a. Polarized neutron beam irradiates the planar waveguides under the grazing angle $\alpha_i$. The sample is

Ta(3 nm)/Ni$_{0.67}$Cu$_{0.33}$(15)/TbCo$_{11}$(150)/
Ni$_{0.67}$Cu$_{0.33}$(50)//Al$_2$O$_3$(substrate)

where the middle layer is the magnetic film TbCo$_{11}$. The external layers are the material Ni(67 at. %)Cu(33 at. %) which is nonmagnetic at room temperature. SLD is presented in Fig. 4a. For spin up and down SLD of the magnetic film is slightly different. Inside the middle layer, the neutron wave function density $\left| \psi(z) \right|^2$ is resonantly enhanced in the direction $z$ perpendicular to the layers. According to the theory of neutron resonances in planar waveguides [17] the enhanced neutron standing waves are formed in the guiding layer at the periodic conditions for the neutron wavefunction phase:

$$\Phi(k_{0z}) = 2k_{2z}d + \arg(r_{21}) + \arg(r_{23}) = 2\pi n \quad (2)$$

where $k_{2z}$ is the $z$-component of the neutron wave vector inside the guiding layer, $k_{0z} = k_0 \sin \alpha_i$ is the $z$-component of the neutron wave vector of the incident beam, $k_0 = 2\pi / \lambda$ is the wave vector of the incident beam, $r_{23}$ is the neutron reflection amplitude from the bottom layer and $r_{21}$ is the neutron reflection amplitude from the upper layer, n=0, 1, 2, ... is the order of the resonance. In Fig. 4b the calculated wavefunction density for spin down is shown vs. the coordinate $z$ and the grazing angle of the incident beam. The neutron wavelength is fixed at 4.26 Å. There are resonances of the orders n=0, 1, 2, 3, 4 in the region of total reflection. In Fig. 4c the calculated reflectivity (right axis) and neutron wavefunction density (left axis) is presented vs.

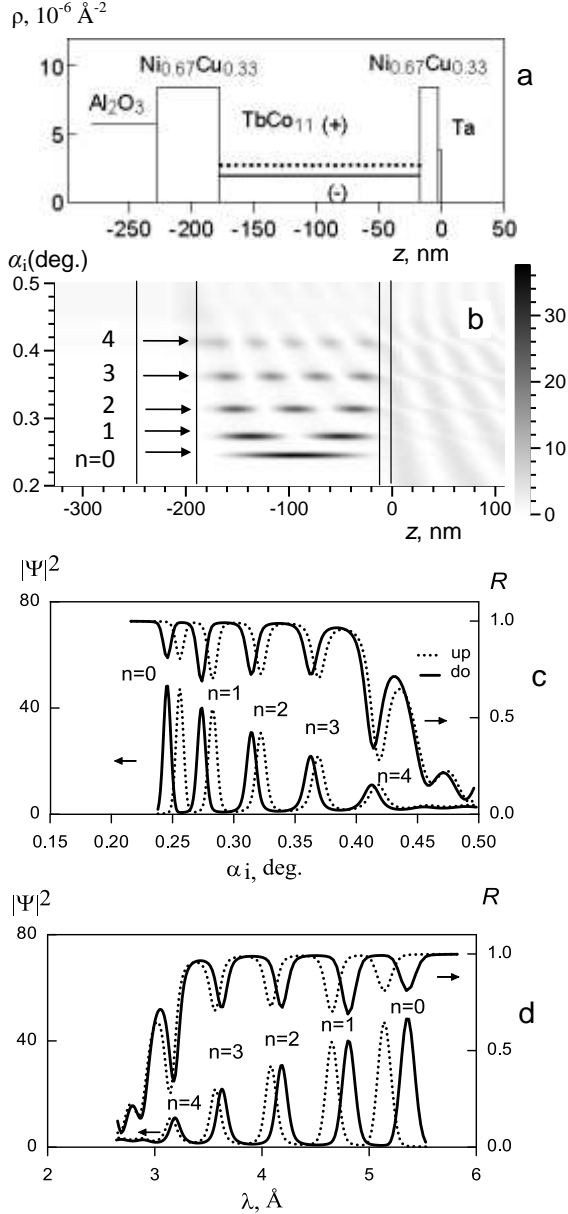

Fig. 4. (a) SLD of the structure
Ta(3 nm)/Ni$_{0.67}$Cu$_{0.33}$ (15)/TbCo$_{11}$ (150)/Ni$_{0.67}$Cu$_{0.33}$(50)//
Al$_2$O$_3$ (substrate). (b) Neutron wavefunction density for spin down vs. the grazing angle of the incident beam and the coordinate z perpendicular to the layers calculated for the neutron wavelength 4.26 Å. (c) Neutron wavefunction density (left axis) and reflectivity (right axis) for spin up (dashed line) and spin down (solid line) vs. the grazing angle of the incident beam calculated for the neutron wavelength 4.26 Å . (d) Neutron wavefunction density (left axis) and reflectivity (right axis) for spin up (dashed line) and spin down (solid line) vs. the neutron wavelength calculated for the grazing angle of the incident beam 0.3°.

the grazing angle of the incident beam for spin up (dashed line) and spin down (solid line) and the magnetization value 500 G. The dips on the reflectivity at total reflection correspond to the maxima of the neutron wavefunction density. The positions of the resonances are shifted for spin up and down and this difference is proportional to the magnetization value. The calculations for various rare-earth containing materials were done in [56]. In Fig. 4d the reflectivity and the neutron wavefunction density for spin up and down are presented vs. the neutron wavelength at the fixed grazing angle of the incident beam 0.3° and the magnetization 500 G. There is also the difference of the resonances positions for spin up and down but the greater order of resonance corresponds to the less neutron wavelength in contrast to the fixed wavelength presentation in Fig. 4c.

The potential energy of neutrons in the magnetic film with magnetic induction $B$ is $\pm\mu B$ where $\mu$ is the neutron magnetic moment, the signs "+" and "-" correspond to neutron spin up and down respectively. In air the magnetic induction $B_0$ is equal to the applied magnetic field $H$ and the potential energy of neutrons in the applied magnetic field is $\pm\mu H$. The part of the kinetic energy of neutron movement in $z$-direction is

$$\frac{h^2}{2m}k_{0z}^2 = \frac{h^2}{2m}k_0^2\sin^2\alpha_i .$$

In general case, the vectors of the applied magnetic field and the magnetic induction of the magnetic film can be non-collinear with the angle $\chi$ between them. On the boundary of two magnetically non-collinear media neutron spin-flip takes place. The spin-flip probability $W$ is depends on the angle $\chi$ as $W \sim \sin^2\chi$. When two vectors **H** and **B** are perpendicular to each other, the angle is $\chi = 90°$ and spin-flip probability is maximal. When two vectors **H** and **B** are collinear, the angle is $\chi = 0°$ and spin-flip probability is zero. For example, at reflection of neutrons from magnetically non-collinear film there are four reflectivities $R^{++}$, $R^{--}$, $R^{+-}$ and $R^{-+}$. The left index "+" or "-" corresponds to neutron spin of the incident beam. The right index "+" or "-" corresponds to neutron spin of the reflected beam. The reflectivity with spin flip is spin-flip probability $R^{+-}, R^{-+} \sim \sin^2\chi$. For the magnetically collinear film spin-flip probability is zero: $R^{+-} = R^{-+} = 0$. In [57] the phenomenon of depolarization of ultracold neutrons at transmission through magnetic foils was explained and spin-flip probability was calculated. In [58] more detailed analysis of the neutron transmission and reflection at the boundary of magnetically non-collinear films was done.

We can write the energy conservation law for spin transition "++" from the external applied magnetic field into the magnetic film inside the waveguide for the fixed neutron wavelength:

$$k_0^2 \sin^2 \alpha_i^{++} + \frac{2m}{\hbar^2} \mu H = (k_{2z}^{++})^2 + \frac{2m}{\hbar^2} \mu B + \rho_2 \quad (3)$$

where $\hbar$ is the Planck's constant, $\rho_2$ is nuclear SLD of the magnetic film, $k_{2z}^{++}$ and $\alpha_i^{++}$ are defined by the conditions of the resonances (2).
For other spin transition we obtain the following expressions:

$$k_0^2 \sin^2 \alpha_i^{--} - \frac{2m}{\hbar^2} \mu H = (k_{2z}^{--})^2 - \frac{2m}{\hbar^2} \mu B + \rho_2 \quad (4)$$

$$k_0^2 \sin^2 \alpha_i^{+-} + \frac{2m}{\hbar^2} \mu H = (k_{2z}^{+-})^2 - \frac{2m}{\hbar^2} \mu B + \rho_2 \quad (5)$$

$$k_0^2 \sin^2 \alpha_i^{-+} - \frac{2m}{\hbar^2} \mu H = (k_{2z}^{-+})^2 + \frac{2m}{\hbar^2} \mu B + \rho_2 \quad (6)$$

The magnetization of the magnetically collinear film is equal to $M = B - H$. From the condition of the resonance $n = 0$ in (3) we can found:

$$k_{2z}^{++} = \frac{1}{2d} \left[ -\arg(r_{21}^{++}) - \arg(r_{23}^{++}) \right] \quad (7)$$

$$k_{2z}^{--} = \frac{1}{2d} \left[ -\arg(r_{21}^{--}) - \arg(r_{23}^{--}) \right] \quad (8)$$

$$\Delta k_{res}^2 = \left( k_{2z}^{--} \right)^2 - \left( k_{2z}^{++} \right)^2 = \\ = O\left[ \arg(r_{21}^{--}), \arg(r_{23}^{--}), \arg(r_{21}^{++}), \arg(r_{23}^{++}) \right] \quad (9)$$

From (3), (4) and (9) the magnetization magnitude can be defined for the fixed neutron wavelength:

$$M = \frac{\hbar^2}{4\mu m} \left( \frac{2\pi}{\lambda} \right)^2 \left[ \sin^2 \alpha_{i0}^{++} - \sin^2 \alpha_{i0}^{--} \right] + \\ + O\left[ \arg(r_{21}^{--}), \arg(r_{23}^{--}), \arg(r_{21}^{++}), \arg(r_{23}^{++}) \right] \quad (10)$$

The magnetization magnitude measured by time-of-flight technique is following:

$$M = \frac{(2\pi\hbar)^2 \sin^2 \alpha_i}{4\mu m} \left[ \frac{1}{\left( \lambda_0^{++} \right)^2} - \frac{1}{\left( \lambda_0^{--} \right)^2} \right] + \\ + O\left[ \arg(r_{21}^{--}), \arg(r_{23}^{--}), \arg(r_{21}^{++}), \arg(r_{23}^{++}) \right] \quad (11)$$

It is shown in [8] that we can neglect by the value $O\left[ \arg(r_{21}^{--}), \arg(r_{23}^{--}), \arg(r_{21}^{++}), \arg(r_{23}^{++}) \right]$.

Thus, measuring experimentally the polarized neutron microbeam intensities "++" and "--" for the resonance $n = 0$ as a function of the incident grazing angle or the neutron wavelength it is possible to extract directly the magnetization value of the weakly magnetic film.

## III. EXPERIMENT

### A. Fixed wavelength mode

Experiment was done at the polarized neutron reflectometer NREX with horizontal sample plane. The fixed neutron wavelength is $\lambda = 4.26$ Å ($\delta\lambda/\lambda = 1.5$ %). The sample was Ta(3 nm)/Ni$_{0.67}$Cu$_{0.33}$(15)/TbCo$_{11}$(150)/Ni$_{0.67}$Cu$_{0.33}$(50)// Al$_2$O$_3$(substrate) with the Al$_2$O$_3$ substrate sizes 25×25×1 mm$^3$. The upper layer Ta was deposited to prevent the film surface from oxidation. The first slit after the monochromator was 0.5 mm of the width and the second slit before the sample was 0.5 mm of the width. The first and the second adiabatic radio-frequency spin-flippers had the efficiency close to 100 %. The polarizing efficiency of the polarizer and the analyzer was 98.8 %. The polarizer and the analyzer are single supermirrors working in transmission mode. The $^3$He two-dimensional position-sensitive detector with spatial resolution 3 mm was used.

To characterize the TbCo$_{11}$ magnetic film, we measured the reflectivities for spin up and down at different applied magnetic field. The procedure of the sample magnetization was following: the film was magnetized in the high negative magnetic field about -10 kOe, then magnetic field was reduced down to zero and then the switched positive magnetic field about +20 Oe was applied. In the field about +550 Oe the film was totally demagnetized. In the experiment we used the analyzer only once to check spin-flip reflectivities in the demagnetized state. The other measurements were done without the analyzer. In Fig. 5a the reflectivities vs. the grazing angle of the incident beam were measured in the applied field 368 Oe. Closed circles, open circles and closed rhombi correspond to spin up, spin down and spin asymmetry $SA = \left( R^{++} - R^{--} \right) / \left( R^{++} + R^{--} \right)$ respectively. There are minima of the resonances n=0, 1, 2, 3 on the total reflection plateau. The reflectivities for spin up and down are merged and spin asymmetry is close to zero. This indicates to low magnetization. In the high magnetic field 4618 Oe (Fig. 5b) there is splitting of the reflectivities. The resonances move to the right for spin up and to the left for spin down. It means that the magnetization of the sample is high. Spin asymmetry is proportional to the magnetization magnitude. Spin asymmetry vs. the applied increasing (closed symbols) and decreasing (open symbols) magnetic field is presented in Figs. 5c,d. In total interval (Fig. 5c) the spin asymmetry is saturated in the high field about 10 kOe. Spin asymmetry for forward and backward directions is different for the field below 8 kOe. In zoom plot in Fig. 5d the magnetization in forward direction is zero at the field about 600 Oe. The arrow indicates the coercive field $H_c$.

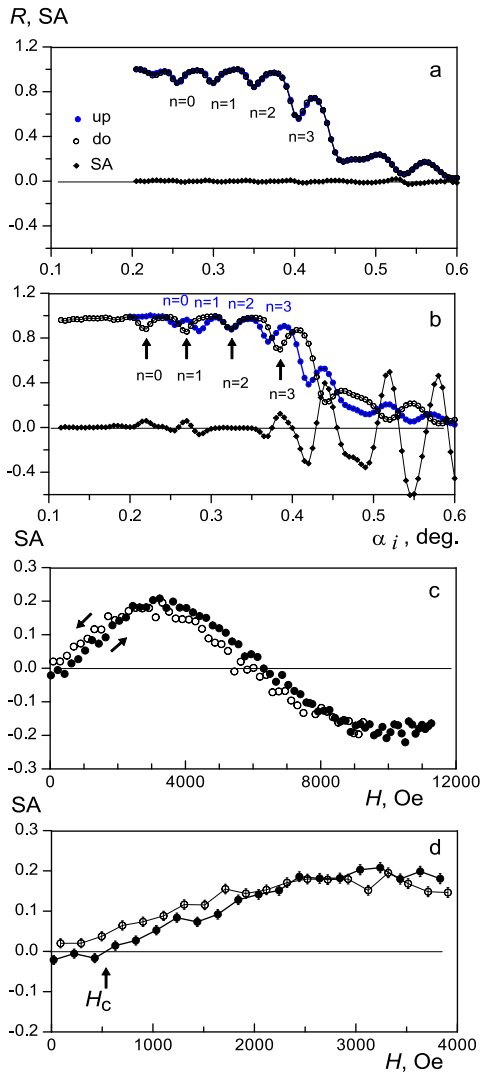

Fig. 5. Specular reflectivity R for spin down (open circles) and up (closed circles) and spin-asymmetry SA (rhombi) vs. the grazing angle of the incident beam for the applied magnetic field 386 Oe (a) and 4618 Oe (b). Spin-asymmetry at the grazing angle of the incident beam 0.415° vs. the applied increasing (closed symbols) and decreasing (open circles) magnetic field plotted in total interval (c) and in zoomed scale (d).

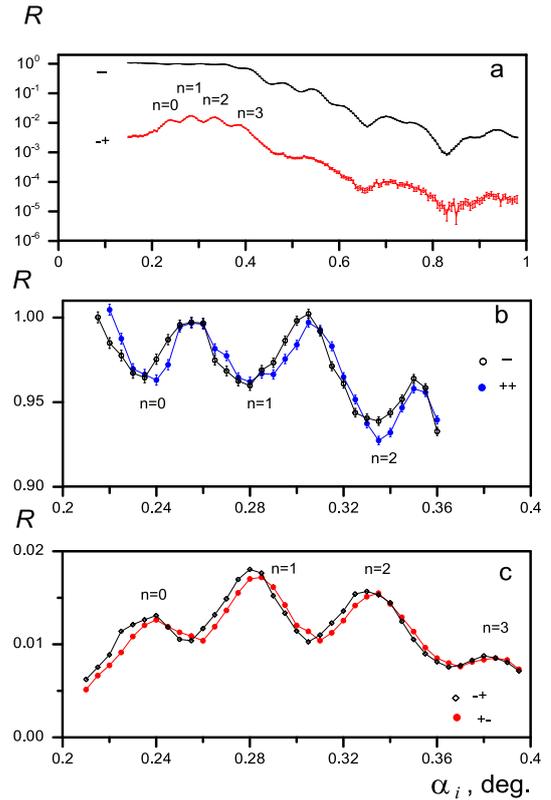

Fig. 6. (a) Specular reflectivity "--" and "-+" for the applied magnetic field 560 Oe. (b) Specular reflectivity "--" and "++" in total reflection region. (c) Specular reflectivity "-+" and "+-" in total reflection region.

The demagnetized film was investigated at the applied field 560 Oe. The analyzer was used. Four reflectivities $R^{++}, R^{--}, R^{+-}, R^{-+}$ were measured using the polarizer and the analyzer. In Fig. 6a the raw (non-corrected) reflectivities $R^{--}$ and $R^{-+}$ are presented. The reflectivities $R^{++}$ and $R^{+-}$ are very similar and are not shown for clarity. The non-corrected reflectivity is the intensity of the specularly reflected beam normalized on the incident angle $\alpha_i$. To extract the net corrected reflectivities from the raw neutron intensities, we should carry out the procedure of the polarization calibration [59-61] taking into account the spin-flippers efficiency and the polarization efficiency of the polarizer and the analyzer. In Fig. 6 the non-corrected spin-flip probability is about 2 %. From fit of the non-corrected reflectivities to the model calculations we obtain that the magnetization about 200 G is directed in the film plane under the angle 10° to the direction of the applied magnetic field. The polarization efficiency of the polarizer and the analyzer is 98.8 % and the imperfection of the polarizing efficiency is 1.2 %. It means that the net corrected spin-flip probability may be estimated as 0.8 %. The fit of the corrected spin-flip reflectivities gives the magnitude of the angle between the magnetization vector and the applied magnetic field about 5°. The splitting of the non spin-flip reflectivities $R^{++}$ and $R^{--}$ corresponds to the magnetization value of the film. The position of the resonance n=0 for $R^{+-}$ coincides with $R^{++}$ and for $R^{-+}$ coincides with $R^{--}$. Thus, from Polarized Neutron Reflectometry we can obtain qualitative information. But quantitative information is indirect and strongly depends on experimental factors. To extract directly the magnetization value we measured the neutron microbeam intensity emitted from the film end face.

In Fig. 7 the two-dimensional map of neutron intensity vs. the grazing angles of the incident and scattered beams is presented for the applied magnetic field 595 Oe (Fig. 7a for spin down and Fig. 7b for spin up) and 2010 Oe (Fig. 7c for spin down and Fig. 7d for spin up). The indices n=0, 1, 2, 3 mark the microbeams of the corresponding resonances (vertical spots). The upper diagonal is the specularly reflected beam and the bottom diagonal is the direct beam suppressed by the beam-stops. The banana-shape curved strong beam is the beam refracted in the substrate. In the low field there is no visible shift of the peak position of the resonance n=0 for spin down and up. In the higher field there is the shift of the peak n=0 for spin down and up.

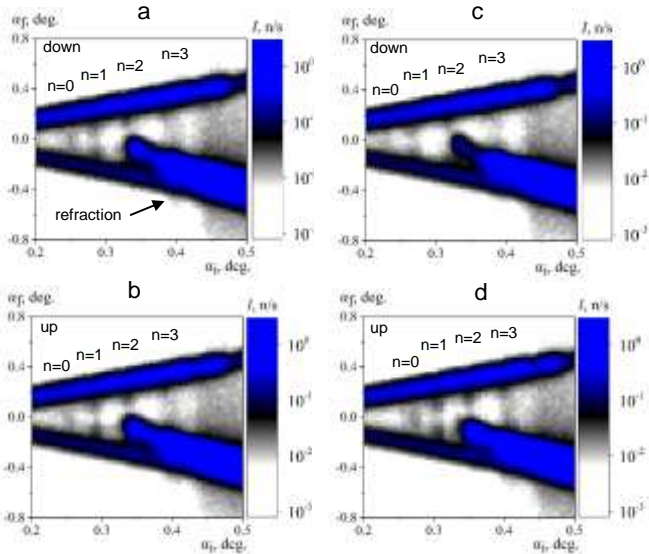

Fig. 7. Neutron intensity vs. the grazing angle of the incident and the scattered beams for the applied field: (a) 595 Oe for spin down; (b) 595 Oe for spin up; (c) 2010 Oe for spin down; (d) 2010 Oe for spin up. The neutron wavelength is 4.26 Å.

In Fig. 8 the neutron microbeam intensity for spin down (open symbols) and spin up (closed symbols) vs. the grazing angle of the incident beam is presented for the applied field in the wide range from 18 to 11400 Oe. The background from the reflected beam increases around $\alpha_i = 0.2°$. In the field 595 Oe the peaks of the resonance n=0 for spin up and down are merged. It means that the magnetization of the films is zero. In lower and higher field there is the splitting of the peaks for spin-up and down with opposite sign. In the high field 11400 Oe the peaks n=0 and n=1 are moved below the angle 0.2°. In this case we can extract magnetization using the angle $\alpha_{i0}^+$ in the high field and the angle $\alpha_{i0} = 0.2505°$ in the demagnetized state with $M$=0 (see Fig. 8d) as

$$M = \frac{h^2}{2\mu m} \left( \frac{2\pi}{\lambda} \right)^2 \left[ \sin^2 \alpha_{i0}^+ - \sin^2 \alpha_{i0} \right] \quad (12)$$

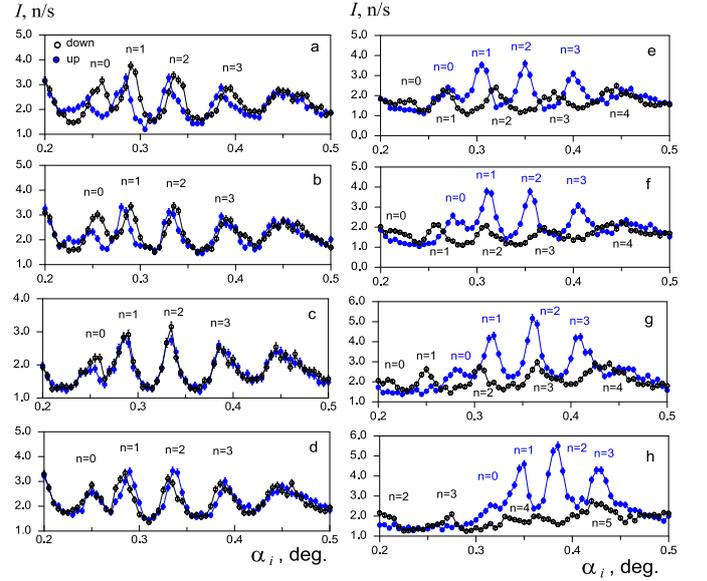

Fig. 8. Neutron microbeam intensity for spin up (closed symbols) and spin down (open symbols) vs. the grazing angle of the incident beam in the applied magnetic field: (a) 18 Oe; (b) 194 Oe; (c) 395 Oe; (d) 595 Oe; (e) 2010 Oe; (f) 3020 Oe; (g) 4007 Oe and (h) 11400 Oe.

In Fig. 9 the neutron microbeam intensity for spin down (open symbols) and spin up (closed symbols) vs. the grazing angle of the incident beam in the applied field around coercive field. The minimal splitting of the peaks for the resonance n=0 is in the field 570 Oe (Fig. 9d). At this scan we reduced the first slit after the monochromator to 0.25 mm, i.e. the angular divergence of the incident beam was 2 times less than for other measurements.

The magnetization value extracted from splitting of the peaks n=0 positions vs. the applied magnetic field is shown in Fig. 10 for total interval (Fig. 10a) and around the coercive filed (Fig. 10b). The data from resonances in Fig. 8 in main features coincides with the hysteresis data in Figs. 5c,d. The saturation is reached in the high field around 10 kOe and the coercive field is about $H_c \approx 565$ Oe. Hysteresis data gives the value of the external field corresponding to saturation, hysteresis and coercive force of the film. And the resonance data give the complementary direct information namely the magnetization value with the experimental sensitivity about 30 G.

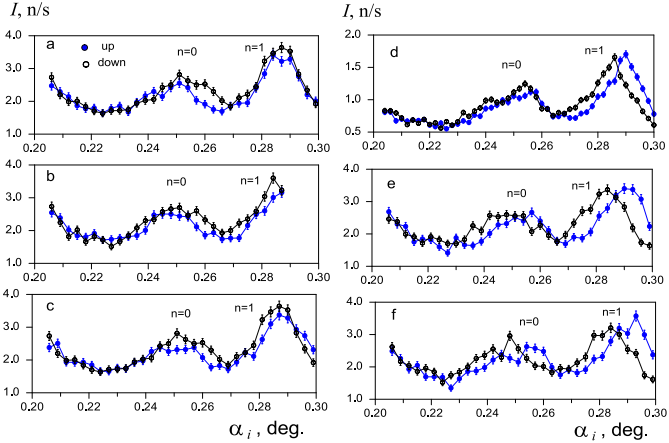

Fig. 9. Neutron microbeam intensity for spin up (closed symbols) and spin down (open symbols) vs. the grazing angle of the incident beam around coercive field: (a) 451 Oe; (b) 530 Oe; (c) 551 Oe; (d) 570 Oe, slit1=0.25 mm; (e) 620 Oe; (f) 640 Oe.

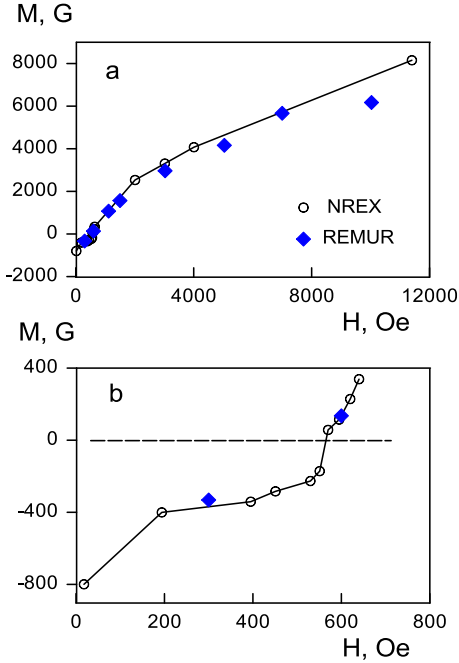

Fig. 10. The magnetization extracted from splitting of the microbeam intensity peaks n=0 vs. the applied magnetic field: (a) total interval; (b) zoom around the coercive field. Open and closed symbols correspond to the results obtained on NREX and REMUR reflectometers respectively.

## B. Fixed wavelength mode

Experiment was done at the polarized neutron time-of-flight reflectometer REMUR [61] at the pulsed reactor IBR-2 (FLNP, JINR, Dubna, Russia). The sample plane is vertical. The polarizer is a single supermirror in reflection geometry. Time-of-flight technique is used. The neutron wavelength resolution 0.02 Å is defined by the reactor pulse width and the time-of-flight base from moderator to detector. The analyzer was not used. Two-coordinate position-sensitive $^3$He detector had the spatial resolution 2.5 mm. The distance sample-detector was 5030 mm and the distance moderator-sample was 29000 mm. The angular divergence of the incident beam was 0.019° and the angular resolution of the detector was 0.028°. In Fig. 11 the neutron spectrum from the combined (thermal and cryogenic) moderator is presented. The neutron intensity drops in 10 times in the interval of neutron wavelengths from 2 to 4 Å.

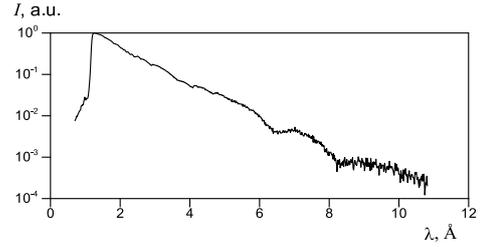

Fig. 11. The neutron spectrum at the REMUR reflectometer from the combined (thermal and cryogenic) moderator of the pulsed reactor IBR-2.

The sample was the same as at the NREX reflectometer. The procedure of the sample magnetization was following. The external magnetic field +10.3 kOe was applied parallel to the sample plane and then the field was decreased down to 0. Then the applied field changed the sign and was increased to -10.3 kOe and then decreased to 0. Then the applied field in the interval from 300 Oe to 10.3 kOe was applied to the sample for the measurements.

Neutron intensity is presented vs. the neutron wavelength and the grazing angle of the incident beam for spin down and spin up in Fig. 12a and 12b for the applied magnetic field 1500 Oe. The indices mark the neutron microbeams of the corresponding resonances (vertical spots). The upper horizontal line is the rest of the reflected beam. The bottom line is the rest of the direct beam. The strong intensity spot around 1.6 Å on the direct beam is the refracted beam. The microbeams with spin down and spin up are shifted to longer and shorter wavelengths respectively.

In Fig. 13 the neutron microbeam intensity is shown vs. the neutron wavelength at the different applied magnetic field: (a) 300 Oe; (b) 1500 Oe and (c) 10.3 kOe. The grazing angle of the incident beam is 0.133°. The closed and open symbols correspond to spin up and down respectively. In the low field 300 Oe the peak splitting is small and corresponds to the negative magnetization. In the field 1500 Oe the splitting is larger. In the high field 10.3 kOe the peaks for the resonances n=0, 1 for spin down are moved to very large neutron wavelengths were the neutron intensity is too low and we cannot see these peaks.

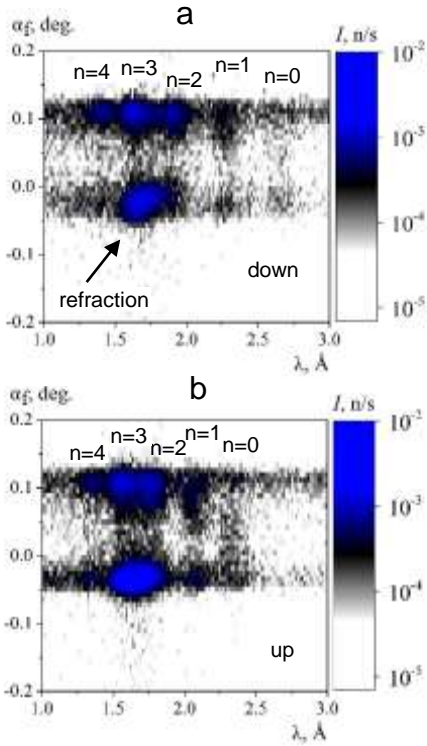

Fig. 12. The neutron intensity at the grazing angle of the incident beam 0.133° vs. the neutron wavelength and the grazing angle of the scattered beam for spin down (a) and up (b). The applied field is 1500 Oe.

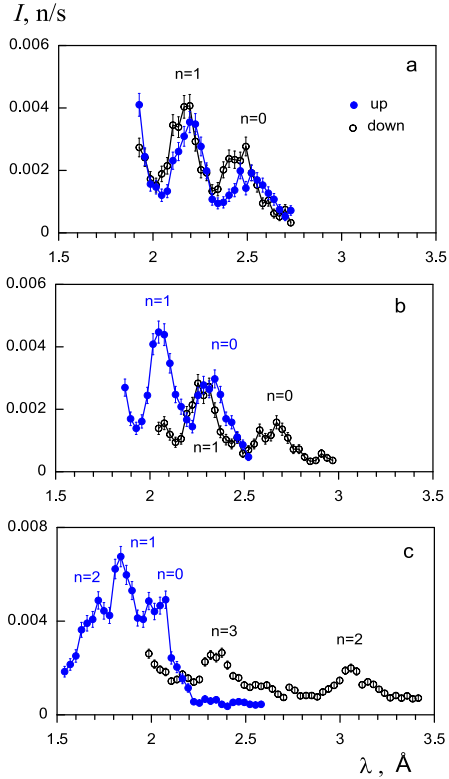

Fig. 13. The neutron microbeam intensity for spin up (closed symbols) and spin down (open symbols) vs. the neutron wavelength at the different applied magnetic field: (a) 300 Oe; (b) 1500 Oe; (c) 10.3 kOe. The grazing angle of the incident beam is 0.133°.

In Fig. 14 the neutron microbeam intensity is shown vs. the neutron wavelength at the different applied magnetic field: (a) 600 Oe; (b) 1114 Oe; (c) 3030 Oe; (d) 5039 Oe; (e) 7001 Oe for spin up; (f) 7001 Oe for spin down. The grazing angle of the incident beam is 0.145°. The closed and open symbols correspond to spin up and down respectively. In the field 600 Oe the magnetization is low with the positive sign. At increasing the field the peak of the resonance $n=0$ moves to the shorter wavelengths for spin up and to the larger wavelengths for spin down. In the field 7001 Oe we cannot see the peak of the resonance $n=0$ for spin down because of the low neutron intensity for large neutron wavelengths.

In Fig. 10 the magnetization of the $TbCo_{11}$ film vs. the applied magnetic field is presented for the NREX (open symbols) and the REMUR (closed symbols) data in total interval (a) and around the coercive field (b). The data obtained on both reflectometers coincide to each other.

In Fig. 15 reflectivities for spin up (closed circles) and down (open circles) and spin asymmetry (rhombi) are presented vs. neutron wavelength for the different applied magnetic field: (a) 600 Oe; (b) 1114 Oe; (c) 3030 Oe and (d) 7001 Oe. The grazing angle of the incident beam is 0.397°. The left axis is reflectivity $R^-$ and spin asymmetry and the right axis is reflectivity $R^+$. In the low field 600 Oe the magnetization is close to zero and spin asymmetry is low. At the total reflection plateau there are the minima of the resonances $n=0, 1, 2, 3$ of the small depth about 0.2 from the total reflection level 1. Therefore spin asymmetry is not large even for high magnetic field. But in the region below the total reflection plateau spin asymmetry is large enough to extract the magnetization value from reflectivity fit.

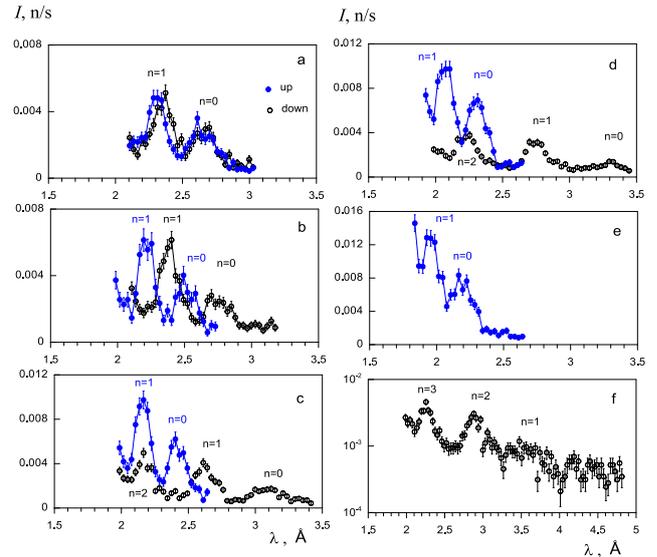

Fig. 14. The neutron microbeam intensity for spin up (closed symbols) and spin down (open symbols) vs. the neutron wavelength at the different applied magnetic field: (a) 600 Oe; (b) 1114 Oe; (c) 3030 Oe; (d) 5039 Oe; (e) 7001 Oe, spin up; (f) 7001 Oe, spin down. The grazing angle of the incident beam is 0.145°.

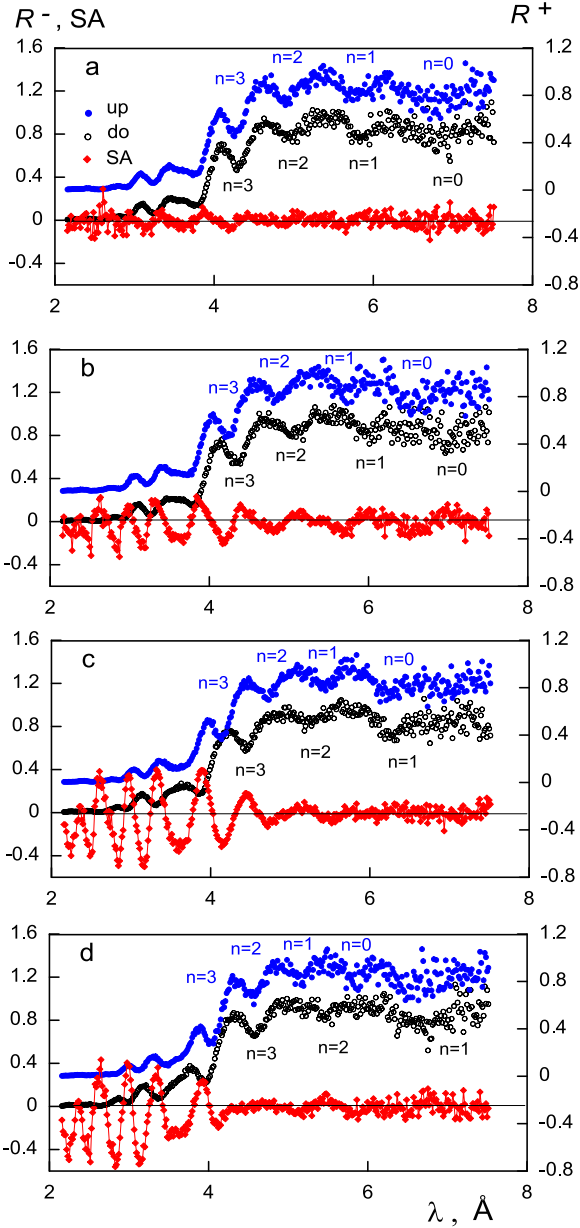

Fig. 15. The reflectivities for spin up (closed circles) and down (open circles) and the spin asymmetry (rhombi) are presented vs. neutron wavelength for the different applied magnetic field: (a) 600 Oe; (b) 1114 Oe; (c) 3030 Oe and (d) 7001 Oe. The grazing angle of the incident beam is 0.397°.

In summary, we measured the magnetization curve of the $TbCo_{11}$ film using fixed wavelength and time-of-flight techniques. Both methods give similar results.

## IV. DISCUSSION

Let us estimate the sensitivity of polarized neutron channeling method in the fixed wavelength mode. From Fig. 9 we can distinguish the shift of two peaks for spin up and down on the magnitude $\delta\alpha_i = 0.0005°$ where $\delta\alpha_i = \alpha^+_i - \alpha^-_i$. From (10) this value corresponds to the magnetization value:

$$\delta M \approx \frac{\hbar^2}{2\mu m}\left(\frac{2\pi}{\lambda}\right)^2 \alpha_i \delta\alpha_i \qquad (13)$$

For $\alpha_i = 0.25°$ and $\lambda = 4.26$ Å we obtain $\delta M = 28$ G.

In the time-of-flight mode the minimal difference of neutron wavelength for spin up and down corresponds to the neutron wavelength resolution $\delta\lambda = 0.02$ Å due to the reactor pulse width. From (11) we can estimate the magnetization sensitivity as

$$M = \frac{(2\pi\hbar)^2 \sin^2 \alpha_i}{2\mu m\lambda^2} \frac{\delta\lambda}{\lambda} \qquad (14)$$

In Fig. 14a, for $\alpha_i = 0.145°$, $\lambda_0 = 2.3$ Å and $\delta\lambda = 0.02$ Å we have $\delta M = 142$ G. We can change the neutron wavelength for the resonance by changing the angle as $\frac{\sin^2 \alpha_i}{\lambda^2} = const$. Then for $\lambda_0 = 4.6$ Å it should be $\delta M = 71$ G and for $\lambda_0 = 6.9$ Å we have $\delta M = 47$ G. The sensitivity is better for a larger wavelength.

In [9,62] other neutron methods for direct investigation of magnetic films are discussed. In the review [9] also the results on polarized neutron channeling in the $TbCo_5$ film can be found. *Larmor precession* of neutron spin at transmission is used for the determination of the magnitude and direction of the magnetic induction about 1 T averaged on the thickness of a magnetic film about 10 µm. Time-of-flight method is used therefore the interval of achievable induction values depends on the interval of neutron wavelength. The resolution of this method is $\delta B / B = \delta\lambda / \lambda$. *Zeeman spatial beam-splitting* can be used for the direct extraction of the magnetic induction about 1 T near the interfaces of two magnetically non-collinear media even in depth or in the domain structures. The accuracy of this method depends on the angular resolution and consists of about 10 %. *Neutron spin resonance* in matter is happened at reflection from magnetic film placed in permanent and oscillating magnetic fields. The frequency of oscillating magnetic field at the resonance corresponds to the frequency of Larmor precession of neutron spin in one domain. Measuring the frequency of the oscillating external field we can directly extract the magnetic induction value in the single domain even in average demagnetized film. The defined accuracy of this method is 4.6 %. The typical value of the magnetic induction defined by these methods is about 0.5 T with accuracy about 10 %. Polarized neutron channeling reported in this communication is a resonant method for weakly magnetic films with magnetization about 100 G or corresponding magnetic induction 0.01 T. Thus, it is a complementary method to other direct neutron methods.

The interference filter also can be used for the determination of the low magnetization. In Fig. 16 the calculations was done for the structure $Ni_{0.67}Cu_{0.33}$ (10nm)/$TbCo_{11}$ (50)/$Ni_{0.67}Cu_{0.33}$ (10)//Si (subst rate) for the neutron wavelength 4.26 Å and the magnetization value 500 G. The magnetic film $TbCo_{11}$ used as a middle layer (Fig. 16a). In reflection there is a resonant dip n=0 shifted for spin up (dashed line) and down (solid line). In transmission there is the corresponding maximum. The width and the depth (or maximum) of the resonance depend on the thickness of the upper and the bottom layers, quality of the layered structures (interface roughness, etc.), experimental resolution. For transmission it is necessary to extract the refracted beam from the direct beam which is a parasitic background. The intensity of the reflected and the refracted beams is in 100 times higher than the microbeam intensity emitted from the end face. Therefore it is possible to use less measuring time or register smaller effects like as spin-flip. From this point of view the interference filter has an advantage in comparison with the planar waveguide. The higher sensitivity of tri-layer resonant structures to the low magnetization value means that for small magnetization the splitting of the peak of the resonance n=0 for spin up and down is greater. It seems that it is possible to optimize the parameters of tri-layer resonant systems (thicknesses of layers, potential well depth, SLD values, etc.) to obtain the highest sensitivity to low magnetization. But it is the subject of more detailed investigation. The final result depends also on experimental conditions. Therefore for correct comparison of the planar waveguide and the interference filter the experiment with the interference filter should be carried out in the future.

## V. CONCLUSIONS

We have demonstrated a new method for the investigation of weakly magnetic films. Intensity of polarized neutrons channeled in the magnetic film inside planar waveguides and emitted from the end face was registered as a function of the grazing angle of the incident beam. From the difference of the microbeam peaks positions for spin up and down we can directly extract the magnetization value of the film. This direct resonant method with high sensitivity about 30 G is complementary for model-dependent polarized neutron reflectometry. We hope that polarized neutron channeling will be demanded for characterization of weakly magnetic films containing rare-earth elements used for the development of magnetic recording and switching.

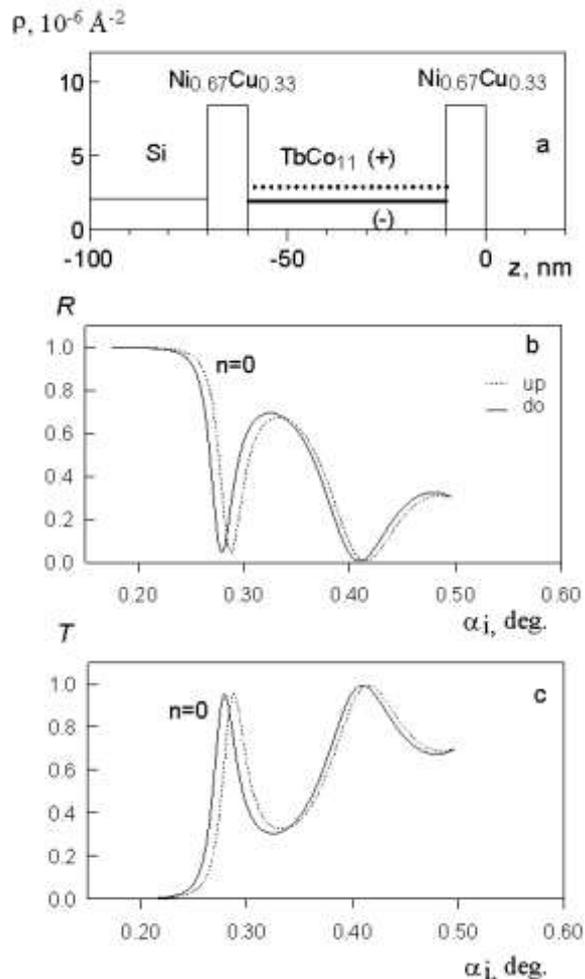

Fig. 16. Calculations for the interference filter $Ni_{0.67}Cu_{0.33}$ (10nm)/$TbCo_{11}$ (50)/$Ni_{0.67}Cu_{0.33}$ (10)//Si (substrate) for the neutron wavelength 4.26 Å. (a) SLD vs. coordinate z perpendicular to the sample surface. (b) Reflectivity for spin up (dashed line) and spin down (solid line). (c) Transmission coefficient for spin up (dashed line) and spin down (solid line).

### Acknowledgements

The authors are thankful to A.V. Petrenko for the technical help at the REMUR reflectometer and V.L. Aksenov and Yu.V. Nikitenko for the interest to this investigation. This work is supported by JINR-Romania scientific project No. 323/21.05.2018, items 89 and 90.